\documentclass{ws-procs975x65}            % default, citations in superscript
\allowdisplaybreaks
\begin{document}
\title{Origin of Temperature of Quark-Gluon Plasma in Heavy Ion Collisions}

\author{Xiao-Ming Xu$^*$}

\address{Department of Physics, Shanghai University,\\
Baoshan, Shanghai 200444, China\\
$^*$E-mail: xmxu@mail.shu.edu.cn}

\begin{abstract}
Initially produced quark-gluon matter at RHIC and LHC does not have a 
temperature. A quark-gluon plasma has a high temperature. From this
quark-gluon matter to the quark-gluon plasma is the early
thermalization or the rapid creation of temperature. Elastic three-parton
scattering plays a key role in the process. The temperature originates from the
two-parton scattering, the three-parton scattering, the four-parton
scattering and so forth in quark-gluon matter.
\end{abstract}


\bodymatter

\bigskip

\noindent Temperature is common in daily life, and people believe that 
temperature exists
no matter how cold matter is. In gas and liquid temperature reflects
the existence of thermal states where all particles randomly move and 
obey isotropic momentum distributions. The average momentum that a massless
particle possesses is proportional to temperature. When temperature equals
zero, the average momentum is zero, and vice versa.
However, there is one system where the average momentum is large but particles
do not obey isotropic momentum distributions, i.e., the system
does not have temperature. The system is
quark-gluon matter that was produced in initial gold-gold nuclear collisions
performed at the Brookhaven National Laboratory Relativistic Heavy Ion
Collider (RHIC) in the last fourteen years. In the parton system the parton
distribution functions in the direction of the incoming nucleus beam
are much larger than those in the direction perpendicular
to the beam direction. The distributions are anisotropic in momentum space, and
do not correspond to a thermal state. Now we must wonder if this quark-gluon
matter always has no temperature. The answer is no, and this matter is exactly
a place for exploring the origin of temperature. 

The reaction $gg \to gg$ was considered by Shuryak \cite{shuryak} to cause
gluon matter to thermalize. He assumed that every gluon suffers from scattering
once in the same period of time. In order to guarantee validity of the
assumption, the cross section for the gluon-gluon scattering must be very large
(empirically be larger than 40 mb). This is not achievable! Even though a
thermalization time of the order of 0.3 fm/$c$ was obtained, the assumption
makes the thermalization time unreliable. To explore the kinetic equilibration
of gluon momenta, inelastic scattering $gg \to ggg$ was taken into account in
Refs. \cite{SM,XG}. $gg \to gg$ and $gg \to ggg$ lead to a gluon-matter
thermalization time larger than 1 fm/$c$. Overpopulation is found to be a key
feature of the Glasma (far-from-equilibrium gluon matter), and can strengthen
the kinetic equilibration. But solutions of a transport equation with the 
elastic $gg$ scattering show that the thermalization time of the Glasma
is also larger than 1 fm/$c$ \cite{BGLMV}.

However, the kinetic equilibration, i.e., the establishment of a thermal
state takes a time less than 1 fm/$c$, as concluded from the elliptic flow data
of hadrons produced in Au+Au collisions at $\sqrt {s_{NN}}=130$ GeV at
RHIC \cite{STAR} and the corresponding explanation
of hydrodynamic calculations which assume early thermalization of quark-gluon
matter and ideal relativistic fluid flow of the quark-gluon
plasma \cite{KHHH,TLS}. It was pointed out by Kolb et al. 
that quark-gluon matter begins to have a temperature at 0.6 fm/$c$ from the
moment when quark-gluon matter is created in initial Au+Au collisions.
The early thermalization thus became the first RHIC puzzle \cite{HK}. 
Even for Pb+Pb collisions at  $\sqrt {s_{NN}}=2.76$ TeV at the Large Hadron 
Collider (LHC), a good explanation of the elliptic flow data of hadrons
\cite{noferini} in hydrodynamic calculations requires that quark-gluon matter
begins to have a temperature at 0.6 fm/$c$ \cite{HSS}. Therefore, the early 
thermalization, i.e., the rapid equilibration also exists at LHC.

To explain the early thermalization or to find the origin of temperature of 
quark-gluon plasma, we think about the prominent feature of initially produced
quark-gluon matter: the parton
number density is as high as 32 ${\rm fm}^{-3}$. At such a high number density
the occurrence probability of parton-parton-parton scattering is larger than 
0.2 \cite{XMCW}. The feature
leads to that the three-parton scattering is important in establishing
thermal states of quark-gluon matter, i.e., creating an initial temperature
of quark-gluon plasma \cite{XSCZ,XSYX}.
For example, gluon-gluon-gluon scattering was discovered to dominate the
evolution of gluon matter towards thermal equilibrium \cite{XSCZ}. We draw the
conclusions from transport equations which include the elastic parton-parton
scattering \cite{CS,CKR} and the elastic scattering of 
gluon-gluon-gluon \cite{XSCZ}, 
gluon-gluon-quark, gluon-gluon-antiquark \cite{XSYX}, 
gluon-quark-quark, gluon-quark-antiquark, gluon-antiquark-antiquark \cite{XX}, 
quark-quark-quark\cite{XPW}, quark-quark-antiquark, quark-antiquark-antiquark
\cite{XMCW}, and antiquark-antiquark-antiquark.
The transport equation for gluon matter is
\begin{eqnarray}
\frac {\partial f_{\mathrm{g1}}}{\partial t}
& + & \vec {\mathrm v}_1 \cdot \vec {\nabla}_{\vec {\mathrm r}} f_{\mathrm g1}
= -\frac {1}{2E_1} \int \frac {d^3p_2}{(2\pi)^32E_2}
\frac {d^3p_3}{(2\pi)^32E_3} \frac {d^3p_4}{(2\pi)^32E_4}
(2\pi)^4           \nonumber    \\
& & \times \delta^4(p_1+p_2-p_3-p_4)
\left\{ \frac {g_{\mathrm G}}{2} \mid 
{\cal M}_{{\mathrm{gg}} \to {\mathrm{gg}}} \mid^2
[f_{\mathrm g1}f_{\mathrm g2}(1+f_{\mathrm g3})(1+f_{\mathrm g4})     \right.
         \nonumber    \\
& & -f_{\mathrm g3}f_{\mathrm g4}(1+f_{\mathrm g1})(1+f_{\mathrm g2})] 
+ g_{\mathrm Q} ( \mid {\cal M}_{{\mathrm{gu}} \to {\mathrm{gu}}} \mid^2
+ \mid {\cal M}_{{\mathrm{gd}} \to {\mathrm{gd}}} \mid^2
              \nonumber    \\
& & +\mid {\cal M}_{{\mathrm g}\bar {\mathrm u} \to {\mathrm g}
\bar {\mathrm u}} \mid^2
+ \mid {\cal M}_{{\mathrm g}\bar {\mathrm d} \to {\mathrm g}\bar {\mathrm d}} 
\mid^2 ) [f_{\mathrm g1}f_{\mathrm q2}(1+f_{\mathrm g3})(1-f_{\mathrm q4})
         \nonumber    \\
& & \left.
-f_{\mathrm g3}f_{\mathrm q4}(1+f_{\mathrm g1})(1-f_{\mathrm q2})]    \right\}
         \nonumber    \\
& & -\frac {1}{2E_1} \int \frac {d^3p_2}{(2\pi)^32E_2}
\frac {d^3p_3}{(2\pi)^32E_3} \frac {d^3p_4}{(2\pi)^32E_4}
\frac {d^3p_5}{(2\pi)^32E_5} \frac {d^3p_6}{(2\pi)^32E_6}
         \nonumber    \\
& & \times (2\pi)^4 \delta^4(p_1+p_2+p_3-p_4-p_5-p_6) 
\left\{ \frac {g_{\mathrm G}^2}{12} \mid 
{\cal M}_{{\mathrm{ggg}} \to {\mathrm{ggg}}} \mid^2 
[f_{\mathrm g1}f_{\mathrm g2}f_{\mathrm g3}      \right.
         \nonumber    \\
& & \times (1+f_{\mathrm g4})(1+f_{\mathrm g5})(1+f_{\mathrm g6})-
f_{\mathrm g4}f_{\mathrm g5}f_{\mathrm g6}
(1+f_{\mathrm g1})(1+f_{\mathrm g2})(1+f_{\mathrm g3})]
         \nonumber    \\
& & + \frac {g_{\mathrm G}g_{\mathrm Q}}{2} 
( \mid {\cal M}_{{\mathrm{ggu}} \to {\mathrm{ggu}}} \mid^2
+\mid {\cal M}_{{\mathrm{ggd}} \to {\mathrm{ggd}}} \mid^2
+\mid {\cal M}_{{\mathrm{gg}}\bar {\mathrm u} 
\to {\mathrm{gg}}\bar {\mathrm u}} \mid^2 
+\mid {\cal M}_{{\mathrm{gg}}\bar {\mathrm d} 
\to {\mathrm{gg}}\bar {\mathrm d}} \mid^2 )
         \nonumber    \\
& & \times [f_{\mathrm g1}f_{\mathrm g2}f_{\mathrm q3}
(1+f_{\mathrm g4})(1+f_{\mathrm g5})(1-f_{\mathrm q6})
-f_{\mathrm g4}f_{\mathrm g5}f_{\mathrm q6}
(1+f_{\mathrm g1})(1+f_{\mathrm g2})
         \nonumber    \\
& & \times (1-f_{\mathrm q3})] + g_{\mathrm Q}^2 [\frac {1}{4} 
\mid {\cal M}_{{\mathrm{guu}} \to {\mathrm{guu}}} \mid^2
+\frac {1}{2} ( \mid {\cal M}_{{\mathrm{gud}} \to {\mathrm{gud}}} \mid^2
              + \mid {\cal M}_{{\mathrm{gdu}} \to {\mathrm{gdu}}} \mid^2 )
         \nonumber    \\
& & +\frac {1}{4} \mid {\cal M}_{{\mathrm{gdd}} \to {\mathrm{gdd}}} \mid^2
    + \mid {\cal M}_{{\mathrm{gu}}\bar {\mathrm u} \to 
{\mathrm{gu}}\bar {\mathrm u}} \mid^2
    + \mid {\cal M}_{{\mathrm{gu}}\bar {\mathrm d} \to 
{\mathrm{gu}}\bar {\mathrm d}} \mid^2
    + \mid {\cal M}_{{\mathrm{gd}}\bar {\mathrm u} \to 
{\mathrm{gd}}\bar {\mathrm u}} \mid^2
         \nonumber        \\
& &     + \mid {\cal M}_{{\mathrm{gd}}\bar {\mathrm d} \to 
{\mathrm{gd}}\bar {\mathrm d}} \mid^2
+\frac {1}{4} \mid {\cal M}_{{\mathrm g}\bar {\mathrm u}\bar {\mathrm u}
                  \to {\mathrm g}\bar {\mathrm u}\bar {\mathrm u}} \mid^2
    +\frac {1}{2} ( \mid {\cal M}_{{\mathrm g}\bar {\mathrm u}\bar {\mathrm d}
                    \to {\mathrm g}\bar {\mathrm u}\bar {\mathrm d}} \mid^2 
    + \mid {\cal M}_{{\mathrm g}\bar {\mathrm d}\bar {\mathrm u} 
           \to {\mathrm g}\bar {\mathrm d}\bar {\mathrm u}} \mid^2 )
         \nonumber    \\
& & + \frac {1}{4} \mid {\cal M}_{{\mathrm g}\bar {\mathrm d}\bar {\mathrm d} 
               \to {\mathrm g}\bar {\mathrm d}\bar {\mathrm d}} \mid^2 ]
[f_{\mathrm g1}f_{\mathrm q2}f_{\mathrm q3}
(1+f_{\mathrm g4})(1-f_{\mathrm q5})(1-f_{\mathrm q6})
           -f_{\mathrm g4}f_{\mathrm q5}f_{\mathrm q6}
         \nonumber    \\
& & \times \left. 
(1+f_{\mathrm g1})(1-f_{\mathrm q2})(1-f_{\mathrm q3})] \right\} ,     
         \nonumber    \\
\end{eqnarray}
and the transport equation for up-quark matter is
\begin{eqnarray}
\frac {\partial f_{\mathrm q1}}{\partial t}
& + & \vec {\mathrm v}_1 \cdot \vec {\nabla}_{\vec {\mathrm r}} f_{\mathrm q1}
= -\frac {1}{2E_1} \int \frac {d^3p_2}{(2\pi)^32E_2}
\frac {d^3p_3}{(2\pi)^32E_3} \frac {d^3p_4}{(2\pi)^32E_4} (2\pi)^4 
         \nonumber    \\
& & \times \delta^4(p_1+p_2-p_3-p_4) \left\{ g_{\mathrm G} \mid 
{\cal M}_{{\mathrm{ug}}  \to {\mathrm{ug}}}  \mid^2
[f_{\mathrm q1}f_{\mathrm g2}(1-f_{\mathrm q3})(1+f_{\mathrm g4})  \right.
         \nonumber    \\
& & -f_{\mathrm q3}f_{\mathrm g4}(1-f_{\mathrm q1})(1+f_{\mathrm g2})]
+ g_{\mathrm Q} (\frac {1}{2} \mid {\cal M}_{{\mathrm{uu}} \to {\mathrm{uu}}}
\mid^2 
+ \mid {\cal M}_{{\mathrm{ud}} \to {\mathrm{ud}}} \mid^2 
         \nonumber    \\
& & 
+ \mid {\cal M}_{{\mathrm u}\bar {\mathrm u} \to {\mathrm u}\bar {\mathrm u}} 
\mid^2
+ \mid {\cal M}_{{\mathrm u}\bar {\mathrm d} \to {\mathrm u}\bar {\mathrm d}}
\mid^2 ) [f_{\mathrm q1}f_{\mathrm q2} (1-f_{\mathrm q3})(1-f_{\mathrm q4})
         \nonumber    \\
& & \left.
-f_{\mathrm q3}f_{\mathrm q4}(1-f_{\mathrm q1})(1-f_{\mathrm q2})]   \right\}
         \nonumber    \\
& & -\frac {1}{2E_1} \int \frac {d^3p_2}{(2\pi)^32E_2}
\frac {d^3p_3}{(2\pi)^32E_3} \frac {d^3p_4}{(2\pi)^32E_4}
\frac {d^3p_5}{(2\pi)^32E_5} \frac {d^3p_6}{(2\pi)^32E_6}
         \nonumber    \\
& & \times (2\pi)^4 \delta^4(p_1+p_2+p_3-p_4-p_5-p_6)
\left\{ \frac {g_{\mathrm G}^2}{4} \mid 
{\cal M}_{{\mathrm{ugg}} \to {\mathrm{ugg}}} \mid^2  \right.
         \nonumber    \\
& & \times [f_{\mathrm q1}f_{\mathrm g2}f_{\mathrm g3}
(1-f_{\mathrm q4})(1+f_{\mathrm g5})(1+f_{\mathrm g6})
-f_{\mathrm q4}f_{\mathrm g5}f_{\mathrm g6}
(1-f_{\mathrm q1})(1+f_{\mathrm g2})
          \nonumber     \\
& &  \times (1+f_{\mathrm g3})] +g_{\mathrm Q}g_{\mathrm G} 
( \frac {1}{2} \mid {\cal M}_{{\mathrm{uug}} \to {\mathrm{uug}}} \mid^2
+\mid {\cal M}_{{\mathrm{udg}} \to {\mathrm{udg}}} \mid^2
+\mid {\cal M}_{{\mathrm u}\bar {\mathrm u}{\mathrm g} 
\to {\mathrm u}\bar {\mathrm u}{\mathrm g}} \mid^2
         \nonumber    \\
& &  +\mid {\cal M}_{{\mathrm u}\bar {\mathrm d}{\mathrm g} 
\to {\mathrm u}\bar {\mathrm d}{\mathrm g}} \mid^2 )
[f_{\mathrm q1}f_{\mathrm q2}f_{\mathrm g3}
(1-f_{\mathrm q4})(1-f_{\mathrm q5})(1+f_{\mathrm g6})
-f_{\mathrm q4}f_{\mathrm q5}f_{\mathrm g6}(1-f_{\mathrm q1})
          \nonumber     \\
& &  \times (1-f_{\mathrm q2})(1+f_{\mathrm g3})]
+ g_{\mathrm Q}^2 [\frac {1}{12} 
\mid {\cal M}_{{\mathrm{uuu}} \to {\mathrm{uuu}}} \mid^2
+\frac {1}{4} ( \mid {\cal M}_{{\mathrm{uud}} \to {\mathrm{uud}}} \mid^2
         \nonumber    \\
& &         + \mid {\cal M}_{{\mathrm{udu}} \to {\mathrm{udu}}} \mid^2 )
+\frac {1}{4} \mid {\cal M}_{{\mathrm{udd}} \to {\mathrm{udd}}} \mid^2
+\frac {1}{2} \mid {\cal M}_{{\mathrm{uu}}\bar {\mathrm u} 
\to {\mathrm{uu}}\bar {\mathrm u}} \mid^2
    +\frac {1}{2} \mid {\cal M}_{{\mathrm{uu}}\bar {\mathrm d} 
\to {\mathrm{uu}}\bar {\mathrm d}} \mid^2
         \nonumber        \\
& &         + \mid {\cal M}_{{\mathrm{ud}}\bar {\mathrm u} 
\to {\mathrm{ud}}\bar {\mathrm u}} \mid^2
            + \mid {\cal M}_{{\mathrm{ud}}\bar {\mathrm d} 
\to {\mathrm{ud}}\bar {\mathrm d}} \mid^2
+\frac {1}{4} \mid {\cal M}_{{\mathrm u}\bar {\mathrm u}\bar {\mathrm u}
                  \to {\mathrm u}\bar {\mathrm u}\bar {\mathrm u}} \mid^2
    +\frac {1}{2} ( \mid {\cal M}_{{\mathrm u}\bar {\mathrm u}\bar {\mathrm d}
                    \to {\mathrm u}\bar {\mathrm u}\bar {\mathrm d}} \mid^2
         \nonumber   \\
& &   + \mid {\cal M}_{{\mathrm u}\bar {\mathrm d}\bar {\mathrm u} 
\to {\mathrm u}\bar {\mathrm d}\bar {\mathrm u}} \mid^2 )
+ \frac {1}{4} \mid {\cal M}_{{\mathrm u}\bar {\mathrm d}\bar {\mathrm d} 
\to {\mathrm u}\bar {\mathrm d}\bar {\mathrm d}} \mid^2 ]
[f_{\mathrm q1}f_{\mathrm q2}f_{\mathrm q3}
(1-f_{\mathrm q4})(1-f_{\mathrm q5})(1-f_{\mathrm q6})
         \nonumber   \\
& &   \left. -f_{\mathrm q4}f_{\mathrm q5}f_{\mathrm q6}
(1-f_{\mathrm q1})(1-f_{\mathrm q2})(1-f_{\mathrm q3})] \right\} ,
         \nonumber    \\
\end{eqnarray}
where massless partons have the velocity vector ${\vec {\rm v}}_1$ and the 
position vector $\vec {\rm r}$; $g_{\mathrm G}$ and $g_{\mathrm Q}$ are the 
spin-color degeneracy factors for the gluon and the quark, respectively;
the distribution functions for 
the gluon, the up quark, the down quark, the up antiquark, and the down 
antiquark are denoted by $f_{\rm gi}$, $f_{\rm ui}$, $f_{\rm di}$, 
$f_{{\bar {\rm u}}{\rm i}}$, and $f_{{\bar {\rm d}}{\rm i}}$, respectively;
$f_{\rm ui}=f_{\rm di}=f_{{\bar {\rm u}}{\rm i}}=f_{{\bar {\rm d}}{\rm i}}
=f_{\rm qi}$. ${\cal M}_{{\mathrm{ijk}} \to {\mathrm{ijk}}}$ of which the
subscript, $\rm i$, $\rm j$, or $\rm k$, stands for a gluon, a quark, or an 
antiquark is the amplitude for the elastic parton-parton-parton scattering. 
Squared amplitudes were calculated 
in perturbative QCD with different number of Feynman diagrams for different
types of scattering \cite{xu1}. When the squared four-momenta of gluon and/or
quark propagators approach zero, the divergences of the squared amplitudes
for the elastic two-parton scattering and the elastic three-parton scattering 
are encountered. The divergences are removed while the propagators are 
regularized by a screening mass which is evaluated from the distribution
functions by a formula in Refs. \cite{BMW,BMS,KK}.
It can be proved from the transport equations that quark-gluon matter in 
thermal nonequilibrium eventually arrives at thermal equilibrium \cite{xu2}.
The larger the squared amplitudes, the shorter the thermalization time of 
quark-gluon matter \cite{xu3}.

Every time the elastic two-parton scattering occurs, the two partons' momenta 
change but the scattering is limited to a plane. Due to the energy-momentum
conservation only two of the six momentum components of the two final partons
are free, i.e., angular distributions of the outgoing partons can be defined.
The probability of producing two partons in an incoming parton
direction is larger than the one in the direction perpendicular to the 
incoming parton direction. In initially produced quark-gluon matter more
partons are in the gold-beam direction than in the direction perpendicular to
the beam direction. On the one hand the collision of two partons in the beam 
direction reduces partons in the beam direction and increases partons in the 
direction perpendicular to the beam direction, on the other hand the collision
of two partons 
in the direction perpendicular to the beam direction reduces partons
in the direction perpendicular to the beam direction and increases partons in 
the beam direction. Since more partons are in the beam direction, 
the scattering reduces parton number in the beam direction and
increases parton number in the direction perpendicular
to the beam direction. Parton momentum distributions gradually become isotropic
in such a manner via the elastic two-parton scattering.

In the elastic three-parton scattering the three initial partons are generally 
not
in a plane, so are the three final partons. Every time the elastic three-parton
scattering occurs, the three partons' momenta change. Even restricted by the
energy-momentum conservation, five of the nine momentum
components of the three final
partons are free. That the parton momentum distributions become isotropic is 
thus expected to be easy via the elastic three-parton scattering. 
Feynman diagrams exhibit the following processes: a 
virtual parton breaks into three final partons,
two final partons, or one final parton and a virtual parton; a virtual parton
combines with an initial parton to form a final parton;
a virtual parton scatters with an initial parton to produce three final 
partons, two final
partons, or one final parton and a virtual parton; a virtual parton coming
from the scattering of two initial partons does not stay in the plane of the
two initial partons. Due to these processes the three final partons are 
generally not in a plane.

Eventually, isotropic parton momentum
distributions are established by the three-parton scattering
and the two-parton scattering.
With anisotropic momentum distributions of partons in quark-gluon matter
created in central Au+Au collisions at $\sqrt {s_{NN}}=200$ GeV at $t=0.2$ 
fm/$c$ \cite{WG}, the transport equations yield isotropic momentum 
distributions
of gluons first at $t=0.52$ fm/$c$ and of quarks first at $t=0.86$ fm/$c$. The
thermalization times of gluon matter and quark matter are 0.32 fm/$c$ and 0.66
fm/$c$, respectively. The temperature acquired is not
close to zero, and is much higher than the QCD critical temperature 0.175 GeV.
Therefore, the early thermalization is a consequence of the two-parton 
scattering 
and the three-parton scattering. Since the four-parton scattering, the
five-parton scattering, etc. have certain occurrence probabilities \cite{XMCW},
they also contribute to the early thermalization.
The temperature of the quark-gluon plasma comes from the elastic and inelastic
two-parton scattering, the three-parton scattering, and the scattering of more
partons.

\section*{Acknowledgments}
I thank X.-J. Qiu for fruitful discussions.
This work was supported by the National Natural Science Foundation of China
under Grant No. 11175111.

%\section{References}

\end{document}